\documentclass{ifacconf}

\usepackage{natbib}        % required for bibliography
\usepackage{amsmath, amssymb,amsfonts, lipsum, mathtools, cuted, color,thmtools}
\usepackage{graphicx}% Include this line if your
\usepackage{enumerate}

\allowdisplaybreaks
\begin{document}
	\begin{frontmatter}
		
		\title{Plug-and-Play Secondary Control for Safety of LTI Systems under Attacks} 

%\thanks[footnoteinfo]{Sponsor and financial support acknowledgment
%goes here. Paper titles should be written in uppercase and lowercase
%letters, not all uppercase.}

\author[First]{Yankai Lin} 
\author[First]{Michelle S. Chong} 
\author[First]{Carlos Murguia}

\address[First]{Department of Mechanical Engineering, Eindhoven University of Technology, the Netherlands \\ (e-mail: \{y.lin2,m.s.t.chong,c.g.murguia\}@ tue.nl).}

\begin{abstract}                % Abstract of not more than 250 words.
We consider the problem of controller design for linear time-invariant cyber-physical systems (CPSs) controlled via networks. Specifically, we adopt the set-up that a controller has already been designed to stabilize the plant. However, the closed loop system may be subject to actuator and sensor attacks. We first perform a reachability analysis to see the effect of potential attacks. To further ensure the safety of the states of the system, we choose a subset of sensors that can be locally secured and made free of attacks. Using these limited resources, an extra controller is designed to enhance the safety of the new closed loop. The safety of the system will be characterized by the notion of safe sets. Lyapunov based analysis will be used to derive sufficient conditions that ensure the states always stay in the safe set. The conditions will then be stated as convex optimization problems which can be solved efficiently. Lastly, our theoretical results are illustrated through numerical simulations.
\end{abstract}

\begin{keyword}
Safety, Cyber-physical systems, Control systems security, Linear systems, Linear matrix inequality.
\end{keyword}

\end{frontmatter}
%===============================================================================

\section{Introduction}
%This document is a template for \LaTeXe. If you are reading a paper or
%PDF version of this document, please download the electronic file
%\texttt{ifacconf.tex}. You will also need the class file
%\texttt{ifacconf.cls}. Both files are available on the IFAC web site.

%Please stick to the format defined by the \texttt{ifacconf} class, and
%do not change the margins or the general layout of the paper. It
%is especially important that you do not put any running header/footer
%or page number in the submitted paper.\footnote{
%This is the default for the provided class file.}
%Use \emph{italics} for emphasis; do not underline.

%Page limits may vary from conference to conference. Please observe the 
%page limits of the event for which your paper is intended.
Many industrial systems are now remotely controlled over communication networks which makes the operation and maintenance of these systems more efficient. However, due to the involvement of the communication channel, cyber-physical systems or CPS (systems that integrate the physical process with communication networks) are often subject to potential attacks which can lead to disastrous outcomes. Examples include the StuxNet malware incident and many other incidents shown in \cite{cardenas2008research}. Thus, preventing disastrous outcomes or in other words, the safety of CPS is of significant importance. 

Various cyber attacks have been modelled and analyzed in \cite{teixeira2015secure} using tools from systems theory. A core idea presented in \cite{teixeira2015secure} is that, the amount of resources needed for attackers to launch an attack against a CPS varies according to the complexity of attacks. In most cases, the attacker wants to keep attacking the system while remaining stealthy. This can be accomplished by the adversary via well-designed attacks with sufficient resources. In this case, normal fault detection and isolation strategies may fail to work \cite{dutta2017stealthy}. %There are also results on how to optimally design such stealthy attacks under certain assumptions, see \cite{chen2018optimal} for instance. 
As detailed later in this paper, we work with attack signals with bounded energy to reflect the fact that adversaries might aim to remain stealthy and that attacks tampering with sensing, actuation, and networked data always have limited resources due to physical constraints. 

One way to deal with attacks on CPS is to design a control scheme that can detect the presence of attacks and then mitigate it, see \cite{fawzi2014secure,chong2015observability} for results on linear systems. This approach ensures security using redundancy and typically requires a large number of observers to completely nullify the effect of attacks. In this work, however, we follow a different approach. We aim to ensure \textit{safety} of the system when attacks are present. Namely, by modifying the dynamics of the closed-loop control system we want the states of interest to stay in a set where safe operation of the system can be guaranteed in the presence of stealthy attacks. This idea has been explored in \cite{murguia2020security}, where the volume of the set of states that attacks can induce in the system is used as a security metric. After appropriate modification of system dynamics, reachable sets of the attack signal should be made as small as possible and fully contained in the safe set. As a result, to compromise the safety of the system, an attacker needs to invest more resources. One possible way is to inject attack signals with larger amplitudes which will be used as a measure of safety of the system against potential attacks. For other security measures of CPS in different settings, we refer interested readers to recent works \cite{bai2015security,tang2019linear,teixeira2015secure2,milovsevic2019estimating} and references therein.  

\begin{figure}[h!]
    \centering
    \includegraphics[width=8.5cm]{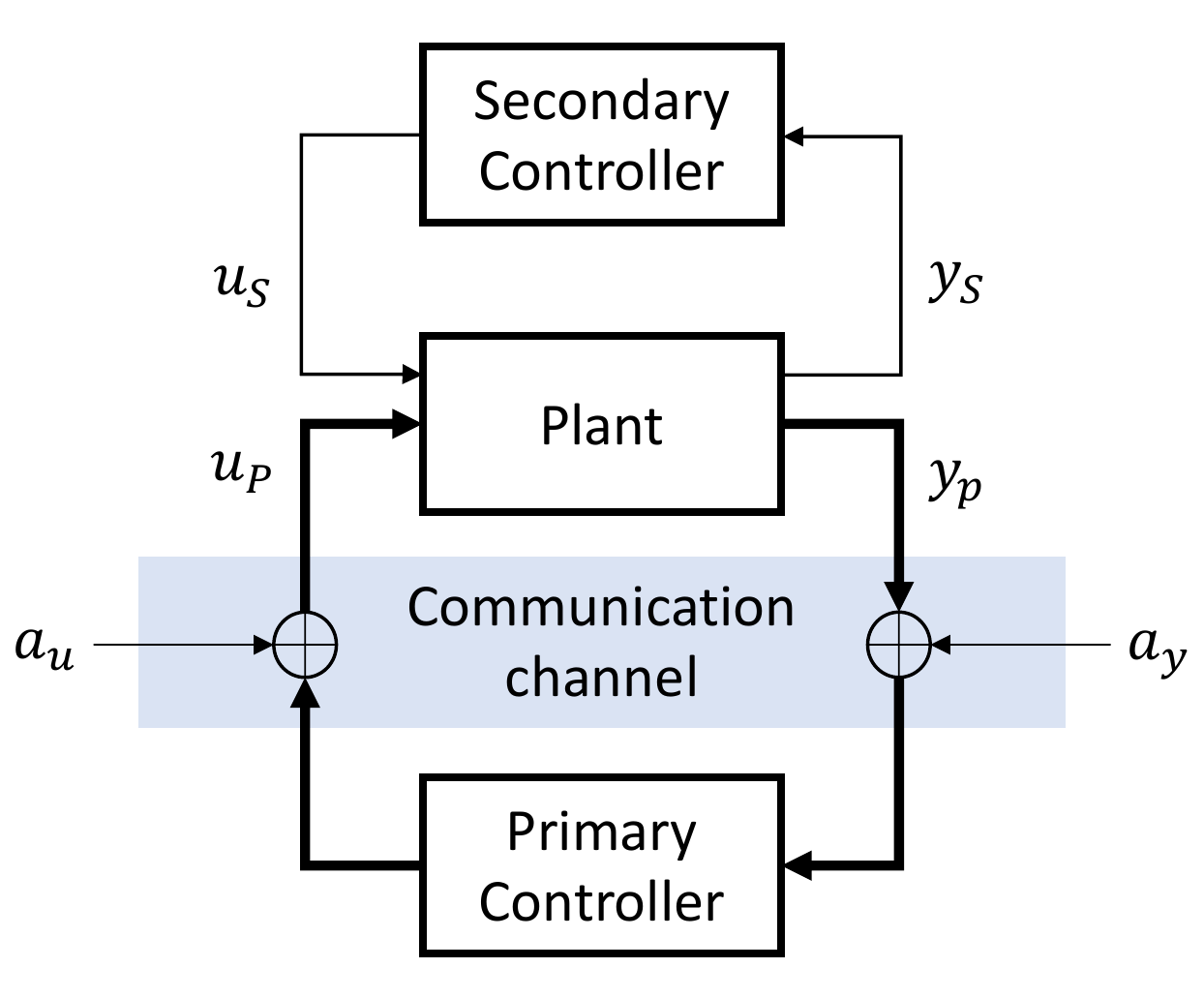}
    \caption{Ensuring safety with a secondary controller.}
    \label{fig:secondary_safety}
\end{figure}

The main contribution of this work is, via set-theoretic methods, a systematic tool to verify the safety of a given linear time-invariant closed-loop control system (a primary controller in connection with the plant) subject to sensor and actuator attacks. In addition, we also provide a tractable method of synthesizing an output feedback linear controller, which we call a secondary controller, that is connected concurrently to the existing closed-loop control system to ensure safety of the overall closed-loop control system. We depict this setup in Figure~\ref{fig:secondary_safety}. This is different from control barrier function based methods for safety-critical systems in \cite{ames2016control}, where state feedback control laws are calculated .

The rest of this manuscript is organized as follows. The problem formulation is given in Section \ref{secPF}. Section \ref{sec3} presents the analysis tool to evaluate the safety of the closed loop system given the pre-designed controller. Section \ref{sec4} proposes optimization-based controller synthesis tools to modify the given system aiming to enhance safety of the overall system. Section \ref{sec5} illustrates the main results via numerical simulations. We will conclude the paper in Section \ref{sec6} and provide some future research directions.

\textit{Notation}: Let $\mathbb{R}$ be the set of real numbers and $\mathbb{R}^{n}$ be the $n$-dimensional Euclidean space. The matrix $I_n$ is used to denote the $n$-dimensional identity matrix and $n$ will be omitted when the dimension is clear. Similarly, $\mathbf{0}$ denotes the zero matrix with appropriate dimensions. For a given vector $x$, For a given square matrix $R$, $\text{Tr}[R]$ denotes the trace of $R$. We use $A\succ 0$ ($A\prec 0$) and $A\succeq 0$ ($A\preceq 0$) to denote the matrix $A$ is positive (negative) definite and positive (negative) semidefinite, respectively.

\section{Problem Formulation}\label{secPF}
We consider the setup shown in Figure \ref{fig:secondary_safety}, where the plant is modelled as a linear time-invariant (LTI) system as follows
\begin{equation}\label{plt}
\begin{split}
    \dot{x}_p&=A_px_p+B_pu\\
    y&=C_px_p,
    \end{split}
\end{equation}
where $x_p\in\mathbb{R}^{n_p}$ is the state vector of the plant, $u\in\mathbb{R}^{n_u}$ and $y\in\mathbb{R}^{n_y}$ are the input and the output of the plant, respectively. We assume that there is a controller which has already been designed to stabilize the plant \eqref{plt},
\begin{equation}\label{control1}
\begin{split}
    \dot{x}_{1}&=A_{1}x_{1}+B_{1}(y+a_y)\\
    u_P&=C_{1}x_{1}+D_{1}(y+a_y)+a_u,
    \end{split}
\end{equation}
which we call the \textit{primary controller} of system (\ref{plt}). It is pre-designed to stabilize (\ref{plt}) with the input signal $u_P\in\mathbb{R}^{n_u}$ and subject to potential cyber attacks denoted by $a=[a_u^\top\ a_y^\top]^\top\in\mathbb{R}^{n_u+n_y}$, where $a_u$ and $a_y$ denote actuator and sensor attacks respectively. Since the primary controller \eqref{control1} is pre-designed without being aware of the attacks, the security and safety of the closed loop may be compromised. Therefore, we propose introducing a \textit{secondary controller}, that runs in conjunction with the primary controller. The secondary controller uses a subset of the sensors and actuators, which are either available locally or known to be safeguarded against malicious manipulation (e.g., via encryption or watermarking). We propose designing a secondary controller which takes the following form
\begin{equation}\label{control2}
\begin{split}
    \dot{x}_{2}&=A_{2}x_{2}+B_{2}y_S\\
    u_S&=C_{2}x_{2}+D_{2}y_S,
    \end{split}
\end{equation}
where $y_S=C_Sy$ is the set of sensors that are available to the secondary controller, and $u_S$ is the control input signal. The overall input is given by
\begin{equation}\label{uuuuu}
    u=u_P+E_uu_S,
\end{equation}
where $E_u$ is the selection matrix that dictates the set of secondary control inputs that will be fed back to the plant \eqref{plt}. Note that the secondary controller uses uncompromised sensor(s) and actuator(s) only, as it is co-located with the plant (i.e., not connected to the network) and we assume that the local sensors and actuators are not subject to network attacks. Consequently, no attack signals appear in (\ref{control2}). The goal of the secondary controller (\ref{control2}) is to ensure that when the overall closed loop system (\ref{plt})-(\ref{uuuuu}) is subject to cyber attacks, the safety of the closed loop can be ensured, i.e., the trajectories of the states of interest of the closed-loop system (\ref{plt})-(\ref{uuuuu}) remain within a given safe set for all $t\geq 0$.

We describe our given safe set by the following ellipsoid $\mathcal{E}(R_{\zeta},\bar{\zeta})$:
\begin{equation}\label{statecons}
    (\zeta-\bar{\zeta})^\top R_{\zeta} (\zeta-\bar{\zeta})\leq 1,
\end{equation}
where $\zeta=[x_p^\top\ x_1^\top\ x_2^\top]^\top:=[\zeta_1^\top\ x_2^\top]^\top$ and $\bar{\zeta}$ is defined similarly. Moreover, we have $R_\zeta\succeq 0$. Ellipsoidal safe set has also been used in existing literature for security and safety, see \cite{romagnoli2020software} for example. If the safe set is of other shapes, (\ref{statecons}) can be used as an outer approximation of the true safe set. Depending on different scenarios, the matrix $R_{\zeta}$ may be rank deficient. In this work, our primary concern are the plant states $x_p$.

We aim to solve the following problems in this paper.
\begin{enumerate}[(Problem I)]
    \item Assess the worst case attacks $a_u$ and $a_y$ that can be tolerated by connecting only the primary controller \eqref{control1} to the plant \eqref{plt}. 
    \item Synthesize the secondary controller (\ref{control2}) such that the attacker needs to invest more resources to violate the safety condition (\ref{statecons}).
\end{enumerate}
We provide a solution to Problem I and II in Section \ref{sec3} and \ref{sec4}, respectively. 

\section{Invariant Set Based Safety Analysis}\label{sec3}
The first step of our analysis is to assess the worst possible attacks that can be tolerated while ensuring safety of the closed loop system. In general, when dealing with cyber attacks, it is natural to assume that the attacks are unbounded. However, depending on the purpose of the attacker, intelligent attackers often seek to remain stealthy and undetected, see \cite{teixeira2015secure}. Thus, we use the following condition of norm bounded attack signals to reflect the fact that the attacker is resource limited and wants to remain undetected, i.e., the attack signals $a$ satisfy the following
\begin{equation}\label{atkcons}
    a^\top R_a a\leq 1,
\end{equation}
with a positive definite matrix $R_a$. Larger valued entries in $R_a$ imply smaller upper bound on the corresponding attack signal $a$.
\begin{rem}
Our analysis is based on invariant set analysis. Theoretically, one should attempt to find the reachable set of the closed loop system subject to attacks and check if the reachable set is a subset of the safe set. However, calculating the exact reachable set of a given system is difficult in general. Thus, we seek an invariant set of the closed loop which can be used as an outer approximation of the reachable set, see \cite{blanchini1999set} and \cite{Escudero22}. 
\end{rem}

The closed loop system consisting of the plant \eqref{plt}, primary controller \eqref{control1}, and the secondary controller \eqref{control2} can be written as 
\begin{equation}\label{closedloop}
    \dot{\zeta}=\mathcal{A}\zeta+\mathcal{B}a,
\end{equation}
where
\begin{equation}\label{mathA}
\begin{split}
    &\mathcal{A}=\\
    &\left[ \begin{array}{ccc} A_p+B_pD_{1}C_p+B_pE_uD_2C_SC_p & B_pC_{p} & B_pE_uC_{2}\\
    B_{1}C_p & A_{1} & \mathbf{0} \\
    B_{2}C_SC_p & \mathbf{0} & A_{2} \end{array} \right],
\end{split}
\end{equation}
\begin{equation}\label{mathB}
    \mathcal{B}=\left[ \begin{array}{cc} B_p & B_pD_{1} \\
    \mathbf{0} & B_1\\
    	\mathbf{0} & \mathbf{0} \end{array} \right].
\end{equation}
Moreover, we define the following partitions,
\begin{equation}
    \mathcal{A}:=\left[ \begin{array}{cc} \mathcal{A}_1 & \mathcal{A}_2\\
    \mathcal{A}_3 & \mathcal{A}_4 \end{array} \right],
\end{equation}
\begin{equation}
    \mathcal{B}:=\left[ \begin{array}{c} \mathcal{B}_1 \\
    \mathcal{B}_2\end{array} \right],
\end{equation}
where
\begin{equation*}
    \mathcal{A}_1=\left[ \begin{array}{cc} A_p+B_pD_{1}C_p+B_pE_uD_2C_SC_p & B_pC_{p} \\
    B_{1}C_p & A_{1}\end{array} \right],
\end{equation*}
\begin{equation*}
    \mathcal{A}_4=A_2,
\end{equation*}
and the expressions of $\mathcal{A}_2$ and $\mathcal{A}_3$ follow from (\ref{mathA}). Similarly, 
\begin{equation*}
    \mathcal{B}_1=\left[ \begin{array}{cc} B_p & B_pD_{1} \\
    \mathbf{0} & B_1\end{array} \right],
\end{equation*}
with $\mathcal{B}_2$ being $\mathbf{0}$.

Since we want to first find the worst attack signals that can be rendered in the safe set by the primary controller (\ref{control1}) only, we set $E_u=\mathbf{0}$ and $C_S=\mathbf{0}$. Moreover, because the secondary controller \eqref{control2} is not connected in feedback with the plant \eqref{plt} in this case, entries corresponding to secondary controller states are set to $\mathbf{0}$. 

We observe that the closed-loop system (\ref{closedloop}) is a LTI system driven by the attack signal $a$. We aim to construct an ellipsoidal outer approximation $\mathcal{E}_{a}$ such that $\mathcal{R}_a\subseteq \mathcal{E}_a$ where $\mathcal{R}_a$ is the reachable set of (\ref{closedloop}). We will solve this problem via Lyapunov analysis and optimization tools.

Consider the quadratic function $V(\zeta)=\zeta^\top P \zeta$ for a positive semi-definite matrix $P\succeq 0$. If we can find a $P$ such that $\dot{V}\leq 0$ whenever $V\geq 1$ along the trajectories of (\ref{closedloop}), then the ellipsoid $\mathcal{E}_P=\{\zeta\ |\ \zeta^\top P\zeta\leq 1\}$ is an invariant set of (\ref{closedloop}) as the states starting inside $\mathcal{E}_P$ cannot leave $\mathcal{E}_P$.

First, for a matrix $Q$ with appropriate dimension, we define the following matrices,
\begin{equation}
    E_1=\left[ \begin{array}{ccc}\mathcal{A}_1^\top Q+Q\mathcal{A}_1 & \mathbf{0} & Q\mathcal{B}_1 \\
    \star & \mathbf{0} & \mathbf{0} \\
    \star &	\star & \mathbf{0} \end{array} \right],
\end{equation}
\begin{equation}
    S=\left[ \begin{array}{ccc}\mathbf{0} & \mathbf{0} & \mathbf{0} \\
    \star & 1 & \mathbf{0} \\
    \star &	\star & -R_a \end{array} \right].
\end{equation}
Since in this part,  the secondary controller has no impact on the closed loop system, it is reasonable to work with only the projection of sets on to the $\zeta_1$-hyperplane. We denote the projection of $\mathcal{E}(R_{\zeta},\bar{\zeta})$ onto the $\zeta_1$-hyperplane by $\hat{\mathcal{E}}(R_{\zeta},\bar{\zeta})$. 

Then, we have the following result.
\begin{prop}\label{prop1}
    Let $E_u=\mathbf{0}$ and $C_S=\mathbf{0}$. Given a positive definite matrix $R_a$, if there exist a positive definite matrix $Q$ and non-negative constants $\alpha,\beta$ satisfying the following inequalities:
    \begin{equation}\label{LMI2}
        -E_1-\alpha F-\beta S\succeq 0,\  s.t. \ \mathcal{E}_Q\subseteq \mathcal{E}(R_{\zeta},\bar{\zeta}),
    \end{equation}
    where $\mathcal{E}_Q=\{\zeta_1\ |\ \zeta_1^\top Q\zeta_1\leq 1\}$, then there exists a forward invariant set for (\ref{closedloop}) under the constraints (\ref{atkcons}) and it is a subset of the safe set $\mathcal{E}(R_{\zeta},\bar{\zeta})$. 
\end{prop}

\begin{pf}
    We consider the vector $[\zeta_1^\top\ 1\  a^\top]^\top$. The condition $\dot{V}(\zeta_1)\leq 0$ can be restated as
\begin{equation}\label{E}
    E_1=\left[ \begin{array}{ccc}\mathcal{A}_1^\top Q+Q\mathcal{A}_1 & \mathbf{0} & Q\mathcal{B}_1 \\
    \star & \mathbf{0} & \mathbf{0} \\
    \star &	\star & \mathbf{0} \end{array} \right]\preceq 0.
\end{equation}
Similarly, we can restate (\ref{atkcons}) as
\begin{equation}\label{S}
    S=\left[ \begin{array}{ccc}\mathbf{0} & \mathbf{0} & \mathbf{0} \\
    \star & 1 & \mathbf{0}\\
    \star &	\star & -R_a \end{array} \right]\succeq 0.
\end{equation}
Finally, the condition $V(\zeta_1)\geq1$ can be written as 
\begin{equation}\label{F}
    F=\left[ \begin{array}{ccc} Q & \mathbf{0} & \mathbf{0} \\
    \star & -1 & \mathbf{0} \\
    \star &	\star & \mathbf{0}\end{array} \right]\succeq 0.
\end{equation}
Since we want to ensure that $\dot{V}(\zeta_1)\leq 0$ holds when (\ref{S}) and (\ref{F}) hold, we can apply the $\mathcal{S}$-procedure (see Section 2.6.3 of \cite{boyd1994linear} and Lemma 1 of \cite{escudero2022analysis}), which states that there should exist non-negative constants $\alpha$ and $\beta$ such that the following holds
\begin{equation*}
    -E_1-\alpha F-\beta S\succeq 0.
\end{equation*}
The extra constraint $\mathcal{E}_Q\subseteq \hat{\mathcal{E}}(R_{\zeta},\bar{\zeta})$ ensures that the resulting invariant set $\mathcal{E}_Q$ is a subset of the safe set $\mathcal{E}(R_{\zeta},\bar{\zeta})$ since the secondary controller states $x_2$ have no impact on the overall system. The invariant set can be constructed as $\{\zeta\ |\ \zeta^\top P \zeta\leq 1\}$ where $Q$ is the projection of $P$ onto the $\zeta_1$-hyperplane.
\hfill $\qed$
\end{pf}

\begin{rem}\label{n-lmi}
    Note that in Proposition \ref{prop1}, the condition (\ref{LMI2}) is in fact not a linear matrix inequality (LMI) since both the matrix $Q$ and the constant $\alpha$ are variables. In practice, one can choose to solve (\ref{LMI2}) for a fixed $\alpha\geq 0$.  
\end{rem}

\begin{rem}
    The result given in Proposition \ref{prop1} is only a sufficient condition for finding an invariant set for (\ref{closedloop}). We are restricting the shape of the invariant set to be an ellipsoid for easier analysis. As a result, if there does not exist a $Q, \alpha$, and $\beta$ that satisfy the conditions in Proposition \ref{prop1}, there may still exist an invariant set for (\ref{closedloop}) which is a subset of the safe set.
\end{rem}

\begin{rem}
    The constraint $\mathcal{E}_Q\subseteq \mathcal{E}(R_{\zeta},\bar{\zeta})$ is a special case of the problem of inner approximation of the intersection of ellipsoids. It is shown in Section 3.7.3 of \cite{boyd1994linear} that via the $\mathcal{S}$-procedure, the problem can be equivalently written as an LMI problem which is convex.
\end{rem}

Proposition \ref{prop1} gives a sufficient condition to check, whether it is possible to keep the states of the closed loop system \eqref{closedloop} 
are always inside the safe set, given the resources which are available to the attacker (characterized by $R_a$). To find out the worst possible attack that can be dealt with by the primary controller alone, one way is to maximize the volume of the ellipsoid induced by (\ref{atkcons}). Since $R_a$ is a positive definite matrix, this is equivalent to minimizing the volume of the ellipsoid induced by $a^\top R_a^{-1} a\leq 1$. It is shown in \cite{kurzhanski1997ellipsoidal} that the volume of the latter ellipsoid is proportional to $\sqrt{\det[R_a]}$, where $\det$ means the determinant of a matrix. The function is again shown to share the same minimizers with the function $\log \det [R_a]$, see Section 3.7 of \cite{boyd1994linear}. Unfortunately, for positive definite $R_a$, $\log \det [R_a]$ is non-convex. This makes us follow the approach used in \cite{murguia2020security} by minimizing a convex upper bound of $\sqrt{\det[R_a]}$.

\begin{lem}[Lemma 4, \cite{murguia2020security}]\label{cvxbound}
    Given any positive definite matrix $R\in\mathbb{R}^{n\times n}$, the following inequality holds
    \begin{equation}
        \det[R]^{\frac{1}{2}}\leq\frac{1}{\sqrt{n}^n}\text{Tr}[R]^{\frac{n}{2}}.
    \end{equation}
    Moreover, $\arg\min [\text{Tr}[R]^{\frac{n}{2}}]=\arg\min[\text{Tr}[R]]$.
\end{lem}
\begin{pf}
    See Lemma 4 of \cite{murguia2020security}. \hfill \qed
\end{pf}

The first result in this paper follows from Proposition \ref{prop1} and is presented below.
\begin{thm}\label{thm1}
    Let $E_u=\mathbf{0}$ and $C_S=\mathbf{0}$. If there exist a positive definite matrix $R_a$, a positive definite matrix $Q$, and non-negative constants $\alpha,\beta$ that solve the following optimization problem:
    \begin{equation}\label{cvxp1}
    \begin{split}
        \underset{Q,R_a,\alpha,\beta}{\min}&\ \text{Tr}[R_a]\\
        s.t. &\ -E_1-\alpha F-\beta S\succeq 0\\
        &\ \mathcal{E}_Q\subseteq \hat{\mathcal{E}}(R_{\zeta},\bar{\zeta}),
    \end{split}
    \end{equation}
    then for all attack signals satisfying (\ref{atkcons}), $\mathcal{E}_P$ is a subset of the safe set $\mathcal{E}(R_{\zeta},\bar{\zeta})$ and is a forward invariant set for (\ref{closedloop}), where $Q$ is the projection of $P$ on to the $\zeta_1$-hyperplane. 
\end{thm}

\begin{rem}
    Note that, in Theorem \ref{thm1}, $R_a$ is also a variable while it is given in Proposition \ref{prop1}. Consequently, to make the optimization problem (\ref{cvxp1}) convex, it should be solved for fixed constants $\alpha$ and $\beta$.  
\end{rem}

\section{Secondary Controller Synthesis}\label{sec4}
We now address the problem of secondary controller synthesis such that the attacker needs to invest more resource to violate the safety condition (\ref{statecons}). The sensor and input selection matrices $C_S$ and $E_u$ are assumed to be pre-selected. That is, the designer first decides which set of sensors are locally available such that they can be secured and which set of secondary controller inputs are to be fed back to the plant. Specifically, given $C_S$ and $E_u$, we want to find $\kappa:=(A_2,B_2,C_2,D_2)$ such that $\text{Tr}[R_a]$ is minimized while there exists an invariant ellipsoid that is contained in the safe set 
represented by (\ref{statecons}). Let us define
\begin{equation}\label{E2}
    E_2=\left[ \begin{array}{ccc}\mathcal{A}^\top P+P\mathcal{A} & \mathbf{0} & P\mathcal{B} \\
    \star & \mathbf{0} & \mathbf{0} \\
    \star &	\star & \mathbf{0} \end{array} \right].
\end{equation}
The synthesis problem can be formulated as follows.
\begin{equation}\label{ncvxp1}
    \begin{split}
        \underset{P,R_a,\alpha,\beta,\kappa}{\min}&\ \text{Tr}[R_a]\\
        s.t. &\ -E_2-\alpha F-\beta S\succeq 0\\
        &\ \mathcal{E}_P\subseteq \mathcal{E}(R_{\zeta},\bar{\zeta}),
    \end{split}
    \end{equation}
Note that, the problem (\ref{ncvxp1}) is non-convex since we have introduced the set of new variables $\kappa$ and there exist products of matrix variables such as $\mathcal{A}^\top P$. This is different from the conditions in Theorem \ref{thm1} where the matrix $\mathcal{A}_1$ is given, making (\ref{E}) linear in optimization variables. Therefore, we follow the approach introduced in \cite{scherer1997multiobjective} to find an invertible change of variables that convexify the problem (\ref{ncvxp1}).

For a positive definite matrix $P$, let it take the following form
\begin{equation}\label{Qchange}
    P=\left[ \begin{array}{cc} Y & N \\
    \star & \tilde{Y}\end{array} \right],\ P^{-1}=\left[ \begin{array}{cc} X & M \\
    \star & \tilde{X}\end{array} \right].
\end{equation}
Let the dimensions of $\mathcal{A}_1$ and $\mathcal{A}_4$ be $n_1$ and $n_2$, respectively. Then we assume $Y$ and $X$ are both in $\mathbb{R}^{n_1\times n_1}$ and $\tilde{Y}$ and $\tilde{X}$ are both in $\mathbb{R}^{n_2\times n_2}$. Moreover, they are positive definite. Via some calculations detailed in \cite{scherer1997multiobjective}, we have the following equality
\begin{equation}\label{pi12}
    P\Pi_1=\Pi_2,
\end{equation}
where
\begin{equation}\label{pi1}
    \Pi_1=\left[ \begin{array}{cc} X & I \\
    M^\top & \mathbf{0}\end{array} \right],\ \Pi_2=\left[ \begin{array}{cc} I & Y \\
    \mathbf{0} & N^\top\end{array} \right].
\end{equation}
To better present how the change of variable works, we define the following matrices
\begin{equation}\label{hatdef}
    \begin{split}
    \hat{A}&=\left[ \begin{array}{cc} A_p+B_pD_{1}C_p & B_pC_{p} \\
    B_{1}C_p & A_{1}\end{array} \right],\ \hat{B}=\left[ \begin{array}{c} B_pE_u \\
    \mathbf{0}\end{array} \right],\\
    &\quad \quad \quad \quad \ \ \ \ \ \ \ \ \hat{C}=\left[ \begin{array}{cc} C_SC_p & \mathbf{0} \end{array} \right].
    \end{split}
\end{equation}
It is obvious that the matrices defined in (\ref{hatdef}) are all given. Moreover, it can be verified that
\begin{equation}\label{connectionschr}
    \mathcal{A}_1=\hat{A}+\hat{B}D_2\hat{C},\ \mathcal{A}_2=\hat{B}C_2, \mathcal{A}_3=B_2\hat{C},\ \mathcal{A}_4=A_2. 
\end{equation}

Then, the change of variables are given as follows
\begin{equation}\label{changeofva}
\begin{split}
\mathbf{A}&=Y\mathcal{A}_1X+Y\mathcal{A}_2M^\top+N\mathcal{A}_3X+N\mathcal{A}_4M^\top\\
\mathbf{B}&=NB_2+Y\hat{B}D_2\\
\mathbf{C}&=C_2M^\top+D_2\hat{C}X\\
\mathbf{D}&=D_2.
\end{split}
\end{equation}

From (\ref{pi1}) and (\ref{changeofva}), after some calculations it can be derived that
\begin{equation}\label{congtran1}
\begin{split}
        \Pi_1^\top P \mathcal{A}\Pi_1=\Pi_2^\top\mathcal{A}\Pi_1&=\left[ \begin{array}{cc} \hat{A}X+\hat{B}\mathbf{C} & \hat{A}+\hat{B}\mathbf{D}\hat{C}\\
    \mathbf{A} & Y\hat{A}+\mathbf{B}\hat{C} \end{array} \right]\\
    :&=A(\eta),
    \end{split}
\end{equation}
\begin{equation}\label{congtran2}
    \Pi_1^\top P\mathcal{B}=\Pi_2^\top \mathcal{B}=\left[\begin{array}{c} \mathcal{B}_1\\
    Y\mathcal{B}_1 \end{array} \right]:=B(\eta).
\end{equation}
Note that, (\ref{congtran1}) and (\ref{congtran2}) are linear in the new variables $\eta:=(X, Y, \mathbf{A}, \mathbf{B}, \mathbf{C},\mathbf{D})$. Based on (\ref{congtran1}) and (\ref{congtran2}), if we perform the congruence transformation with $\text{diag}(\Pi_1,I)$ on (\ref{E2}), we have
\begin{equation}\label{E2bf}
    \mathbf{E}_2=\left[ \begin{array}{ccc}A(\eta)^\top+A(\eta) & \mathbf{0} & B(\eta) \\
    \star & \mathbf{0} & \mathbf{0} \\
    \star &	\star & \mathbf{0} \end{array} \right].
\end{equation}
After applying the same transformation to $F$ and $S$ in (\ref{ncvxp1}), we have
\begin{equation}\label{FSbf}
    \mathbf{F}=\left[ \begin{array}{ccc}P(\eta) & \mathbf{0} & \mathbf{0} \\
    \star & -1 & \mathbf{0} \\
    \star &	\star & \mathbf{0} \end{array} \right],\ \mathbf{S}=S,
\end{equation}
where
\begin{equation}\label{congtran3}
    P(\eta)=\Pi_1^\top P\Pi_1=\left[\begin{array}{cc} X & I\\
    I & Y \end{array} \right].
\end{equation}
Next, we handle the other constraint in (\ref{ncvxp1}). To simplify the analysis, we make the following assumption.
\begin{assum}\label{Ass1}
    Let $Q$ be the projection of $P$ on to the $\zeta_1$-hyperplane, if $\mathcal{E}_Q\subseteq \hat{\mathcal{E}}(R_{\zeta},\bar{\zeta})$ holds, then $\mathcal{E}_P\subseteq \mathcal{E}(R_{\zeta},\bar{\zeta})$ holds.
\end{assum}
Assumption \ref{Ass1} essentially means that the states of the secondary controller $\zeta_2$ is not taken into account by the safe set. We believe this assumption is reasonable, since safety of the states of the plant are often the primary concern in most applications.

Since $P$ has the block structure given in (\ref{Qchange}), from Corollary 2 of \cite{murguia2020security}, it can be verified that $Q=Y-N^\top\tilde{Y}^{-1}N=X^{-1}$ using matrix inversion lemma. From Assumption \ref{Ass1}, the constraint $\mathcal{E}_P\subseteq \mathcal{E}(R_{\zeta},\bar{\zeta})$ can be equivalently written as $\zeta_1^\top Q \zeta_1\leq 1\implies (\zeta_1-\bar{\zeta}_1)^\top \hat{R}_{\zeta} (\zeta_1-\bar{\zeta}_1)\leq 1$, where $\hat{R}_{\zeta}$ is such that 
\begin{equation}\label{stateconsP}
    \hat{\mathcal{E}}(R_{\zeta},\bar{\zeta})=\{\zeta_1\ |\ (\zeta_1-\bar{\zeta}_1)^\top \hat{R}_{\zeta} (\zeta_1-\bar{\zeta}_1)\leq 1\}.
\end{equation}
Since we have $Q=X^{-1}$, the condition $\zeta_1^\top Q \zeta_1\leq 1$ can be equivalently stated as
\begin{equation*}
    [\zeta_1^\top\ 1]\left[ \begin{array}{cc} X^{-1} & \mathbf{0} \\
    \star & -1\end{array} \right][\zeta_1^\top\ 1]^\top\leq 0.
\end{equation*}
Similarly, the condition $\zeta_1^\top\hat{R}_{\zeta}\zeta_1\leq 1$ can be written as 
\begin{equation*}
    [\zeta_1^\top\ 1]\left[ \begin{array}{cc} \hat{R}_{\zeta} & -\hat{R}_{\zeta}\bar{\zeta}_1 \\
    \star & \bar{\zeta}_1^\top\hat{R}_{\zeta}\bar{\zeta}_1-1\end{array} \right][\zeta_1^\top\ 1]^\top\leq 0.
\end{equation*}
Using $\mathcal{S}$-procedure again, $\zeta_1^\top Q \zeta_1\leq 1\implies (\zeta_1-\bar{\zeta}_1)^\top \hat{R}_{\zeta} (\zeta_1-\bar{\zeta}_1)\leq 1$ holds when there exists a non-negative constant $\delta$ such that the following holds
\begin{equation}\label{beforeS}
   \left[ \begin{array}{cc} \hat{R}_{\zeta} & -\hat{R}_{\zeta}\bar{\zeta}_1 \\
    \star & \bar{\zeta}_1^\top\hat{R}_{\zeta}\bar{\zeta}_1-1\end{array} \right]-\delta\left[ \begin{array}{cc} X^{-1} & \mathbf{0} \\
    \star & -1\end{array} \right]\preceq 0.
\end{equation}
Note that (\ref{beforeS}) is nonlinear in the variable $X$. Thus, we apply another congruence transformation with $\text{diag}(X,I)$ to (\ref{beforeS}) leading to
\begin{equation}\label{afterS}
   \left[ \begin{array}{cc} X\hat{R}_{\zeta}X & -X\hat{R}_{\zeta}\bar{\zeta}_1 \\
    \star & \bar{\zeta}_1^\top\hat{R}_{\zeta}\bar{\zeta}_1-1\end{array} \right]-\delta\left[ \begin{array}{cc} X & \mathbf{0} \\
    \star & -1\end{array} \right]\preceq 0.
\end{equation}
Note that the first term of (\ref{afterS}) can be written as
\begin{equation*}
   \left[ \begin{array}{cc} \mathbf{0} & -X\hat{R}_{\zeta}\bar{\zeta}_1 \\
    \star & \bar{\zeta}_1^\top\hat{R}_{\zeta}\bar{\zeta}_1-1\end{array} \right]+[X\  \mathbf{0}]^\top\hat{R}_{\zeta}[X\ \mathbf{0}].
\end{equation*}
Then, via the Schur complement result (see \cite{boyd1994linear}, for example), inequality (\ref{afterS}) is equivalent to
\begin{equation}\label{afterS-Schur}
   \left[ \begin{array}{ccc} \mathbf{0} & -X\hat{R}_{\zeta}\bar{\zeta}_1 & -X\\
    \star & \bar{\zeta}_1^\top\hat{R}_{\zeta}\bar{\zeta}_1-1 & \mathbf{0}\\
    \star & \star & -\hat{R}_{\zeta}^{-1}\end{array} \right]-\delta\left[ \begin{array}{ccc} X & \mathbf{0} & \mathbf{0}\\
    \star & -1 & \mathbf{0}\\
    \star & \star & \mathbf{0} \end{array} \right]\preceq 0,
\end{equation}
which is linear in $X$.

To summarize, the problem (\ref{ncvxp1}) can be equivalently formulated as
\begin{equation}\label{ncvxp1congu}
    \begin{split}
        \underset{\eta,\alpha,\beta,\delta}{\min}&\ \text{Tr}[R_a]\\
        s.t. &\ \mathbf{E}_2+\alpha \mathbf{F}+\beta \mathbf{S}\preceq 0\\
        &\ P(\eta)\succ0\\
        &\ \mathbf{J}-\delta \mathbf{L}\preceq 0,
    \end{split}
    \end{equation}
where
\begin{equation*}
    \mathbf{J}=\left[ \begin{array}{ccc} \mathbf{0} & -X\hat{R}_{\zeta}\bar{\zeta}_1 & -X\\
    \star & \bar{\zeta}_1^\top\hat{R}_{\zeta}\bar{\zeta}_1-1 & \mathbf{0}\\
    \star & \star & -\hat{R}_{\zeta}^{-1}\end{array} \right],\ \mathbf{L}=\left[ \begin{array}{ccc} X & \mathbf{0} & \mathbf{0}\\
    \star & -1 & \mathbf{0}\\
    \star & \star & \mathbf{0} \end{array} \right].
\end{equation*}
Notice that  (\ref{ncvxp1congu}) can be efficiently solved for fixed non-negative constants $\alpha,\beta$ and $\delta$. Since $P(\eta)\succ 0$, we have $Y\succ 0$ and $X-Y^{-1}\succ 0$. Moreover, by the definition of $\Pi_1$ and $\Pi_2$, we have $MN^{\top}=I-XY$ guaranteed to be non-singular. Therefore, once (\ref{ncvxp1congu}) is solved, appropriate $M$ and $N$ can always be found. Lastly, the secondary controller variable $\kappa$ can be found via (\ref{changeofva}) in the order of $D_2$, $C_2$, $B_2$, and $A_2$, see \cite{scherer1997multiobjective}.

We summarize the main result presented above in the following theorem.
\begin{thm}\label{thm2}
    Suppose Assumption \ref{Ass1} holds and $n_1=n_2$. Given $E_u$ and $C_S$. If there exist a positive definite matrix $R_a$, a positive definite matrix $P(\eta)$, and non-negative constants $\alpha,\beta$, and $\delta$ that solve the optimization problem (\ref{ncvxp1congu}), then for all attack signals satisfying (\ref{atkcons}), $\mathcal{E}_P$ is a subset of the safe set $\mathcal{E}(R_{\zeta},\bar{\zeta})$ and is a forward invariant set for (\ref{closedloop}). 
\end{thm}

When $R_a$ is given, we can give a synthesis result similar to Proposition \ref{prop1}.
\begin{cor}\label{cor1}
    Suppose Assumption \ref{Ass1} holds and $n_1=n_2$. Given $R_a$, $E_u$, and $C_S$. If there exists a positive definite matrix $P(\eta)$, and non-negative constants $\alpha,\beta$, and $\delta$ that satisfy
    \begin{equation*}
        \begin{split}
            &\ \mathbf{E}_2+\alpha \mathbf{F}+\beta \mathbf{S}\preceq 0\\
        &\ P(\eta)\succ0\\
        &\ \mathbf{J}-\delta \mathbf{L}\preceq 0,
        \end{split}
    \end{equation*}
    then for all attack signals satisfying (\ref{atkcons}), $\mathcal{E}_P$ is a subset of the safe set $\mathcal{E}(R_{\zeta},\bar{\zeta})$ and is a forward invariant set for (\ref{closedloop}). 
\end{cor}

Corollary \ref{cor1} gives a sufficient condition to verify the possibility of designing an output feedback linear controller to ensure safety of the closed loop system. For example, if the optimization problem in Proposition \ref{prop1} is infeasible meaning the states of the system with primary controller only may be driven outside the safe set. However, if the problem in Corollary \ref{cor1} is feasible under the same $R_a$, then it can be guaranteed that with the addition of the secondary controller, the closed loop system can be rendered safe.

\section{Numerical Case Study}\label{sec5}
We now illustrate the main result of this work via numerical simulations. We will consider a  linear time-invariant plant with two inputs and two outputs  that is stabilized by a linear primary controller. However, under potential attacks that satisfy (\ref{atkcons}), the safety of the overall system can not be verified via Proposition \ref{prop1}. Given the same $R_a$, we attempt to ensure safety of the system via Corollary \ref{cor1} by designing a secondary output feedback linear controller. 

For the secondary controller, we use one secured sensor in local feedback and modify one of the outputs. This makes $E_u=[1\ 0]^\top$ and $C_S=[1\ 0]$. The linear plant is stabilized by the primary controller such that we have
\begin{equation*}
    \hat{A}=\left[ \begin{array}{cc} -1 & 0\\
    0 & -1 \end{array} \right], \hat{B}=\left[ \begin{array}{c} 1\\
    0 \end{array} \right],\hat{C}=[1\ 0].
\end{equation*}
The matrix $R_a$ that characterizes the bound on the attack signals is given by $R_a=\text{diag}(0.5,1,0.4,0.7)$. All inputs and outputs are subject to potential attacks. Lastly, the safe set is given by $\{\zeta\ |\ \zeta_1^\top \hat{R}_{\zeta} \zeta_1\leq 1\}$ with $\hat{R}_{\zeta}=0.022I_2$ which is a sphere centred around the origin.

\begin{figure}[ht]
\begin{center}
\includegraphics[width=8cm]{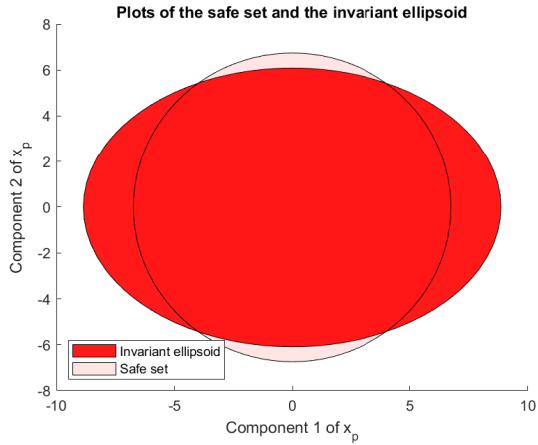}    % The printed column width is 8.4 cm.
\caption{Plots of the safe set and the found invariant set using primary controller only.}
\label{fig:1}
\end{center}
\end{figure}

Via Proposition \ref{prop1}, we can find an invariant set only if we drop the constraint that the invariant set is a subset of the safe set. The plots of the found invariant set and the safe set is shown in Figure \ref{fig:1}. It can be seen that the found invariant set which serves as an outer approximation of the reachable set\footnote{It is mentioned in \cite{murguia2020security} that the approximation is tight for linear systems.} is not a subset of the safe set. This means the system may be driven to an unsafe region by attack signals. 

With the secondary controller design, we aim to guarantee that the invariant ellipsoid is a subset of the safe set. If this is feasible, then there exists at least one controller that keeps the closed loop system safe under all attack signals that satisfy the constraint (\ref{atkcons}). 

We choose set $\alpha=0.25$ and $\delta=0.99$ to make the LMI conditions in Corollary \ref{cor1}. We solve these LMIs to obtain
\begin{equation*}
    X=\left[ \begin{array}{cc} 28.6109 & 0\\
    0 & 31.9965 \end{array} \right], Y=\left[ \begin{array}{cc} 3.9840 & 0\\
    0 & 0.1164 \end{array} \right],
\end{equation*} 
\begin{equation*}
    \mathbf{A}=\left[ \begin{array}{cc} -26.8308 & 0\\
    0 & -0.6765 \end{array} \right],\ \mathbf{B}=\left[ \begin{array}{c} -200.3492\\
    0 \end{array} \right],
\end{equation*}
\begin{equation*}
    \mathbf{C}=[-78.5049\ 0],\ \mathbf{D}=-26.8308.
\end{equation*} 
We then choose $M=I_2$. The fact that $MN^{\top}=I-XY$ yields $N=\left[ \begin{array}{cc} -112.9854 & 0\\
    0 & -2.7259 \end{array} \right]$. The controller parameters can then be solved as follows in the order of $D_2$, $C_2$, $B_2$, and lastly $A_2$ using (\ref{changeofva}).
\begin{equation*}
    A_2=\left[ \begin{array}{cc} -27.2049 & 0\\
    0 & -1.1187 \end{array} \right],\ B_2=\left[ \begin{array}{c} 0.8271\\
    0 \end{array} \right],
\end{equation*}
\begin{equation*}
    C_2=[689.1488\ 0],\ D_2=-26.8308.
\end{equation*}
\begin{figure}[ht]
\begin{center}
\includegraphics[width=8cm]{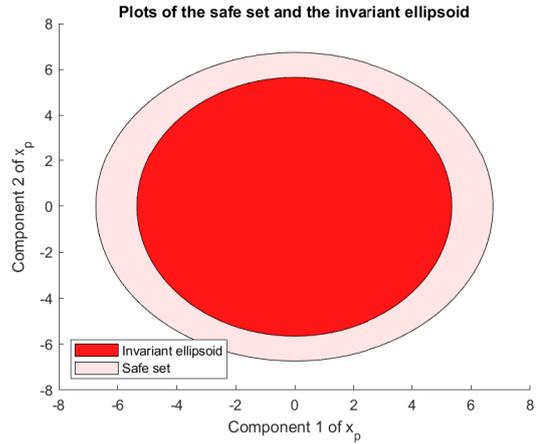}    % The printed column width is 8.4 cm.
\caption{Plots of the safe set and the found invariant set with the addition of the secondary controller.}
\label{fig:2}
\end{center}
\end{figure}

\begin{figure}[ht]
\begin{center}
\includegraphics[width=8cm]{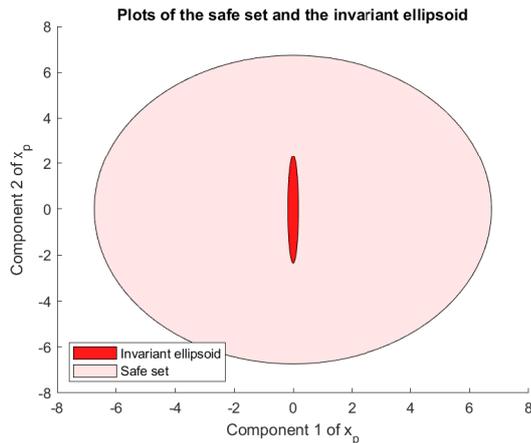}    % The printed column width is 8.4 cm.
\caption{Plots of the safe set and the found invariant set with the addition of the secondary controller that minimizes $\text{Tr}[X]$.}
\label{fig:3}
\end{center}
\end{figure}

It can be seen from Figure \ref{fig:2} that, the found invariant ellipsoid is now a subset of the safe set. Therefore, by adding the secondary controller to the system, we rule out the possibility of unsafe operations of the system subject to attacks satisfying (\ref{atkcons}). One can also try to minimize $\text{Tr}[X]$ under the constraints given in Corollary \ref{cor1}. Recall that via Lemma \ref{cvxbound}, we prove that $\text{Tr}[X]$ can be used as a convex upper bound on the volume of $\mathcal{E}_Q$, with $Q$ being the projection of $P$ on to the $\zeta_1$-hyperplane. The following controller is synthesized if we minimize $\text{Tr}[X]$,
\begin{equation*}
    A_2=\left[ \begin{array}{cc} -1.1745\times 10^8 & 0\\
    0 & -1.8725 \end{array} \right],\ B_2=\left[ \begin{array}{c} 5.0553\times 10^6\\
    0 \end{array} \right],
\end{equation*}
\begin{equation*}
    C_2=[-977.4890\ 0],\ D_2=-232.8737.
\end{equation*}
The plot of the invariant ellipsoid is shown in Figure. \ref{fig:3}. It can be seen that, the volume of the invariant set in Figure. \ref{fig:3} is significantly reduced compared to the one shown in Figure. \ref{fig:2}. 
\section{Conclusions}\label{sec6}
We have used an invariant set based method to provide a framework for checking the safety of LTI systems subject to sensor and actuator attacks by resource limited adversaries. In addition, by using a subset of sensors and secure feedback, a plug-and-play linear secondary controller synthesis problem is also solved to recover the safety of the overall system. The initial synthesis problem is not convex, but can be rendered so with a congruence transformation, which yields an LMI condition that can be solved efficiently. The effectiveness of the proposed design is illustrated via a numerical example, where a secondary controller is designed to guarantee safety which can not be guaranteed by the primary controller alone.

One possible future research direction is the investigation of how to smarly choose the matrices $C_S$ and $E_u$. In our work, we assume it is given, i.e., the designer first chooses which sensors are to be secured and used for secondary controller design and how the output of the secondary controller is fed back to the plant. When the size of the system is large, it is desired to have a more efficient and intelligent approach to choosing which set of sensors are to be used for the secondary controller. 

\bibliography{ref}             % bib file to produce the bibliography
                                                     % with bibtex (preferred)
                                                   
%\begin{thebibliography}{xx}  % you can also add the bibliography by hand

%\bibitem[Able(1956)]{Abl:56}
%B.C. Able.
%\newblock Nucleic acid content of microscope.
%\newblock \emph{Nature}, 135:\penalty0 7--9, 1956.

%\bibitem[Able et~al.(1954)Able, Tagg, and Rush]{AbTaRu:54}
%B.C. Able, R.A. Tagg, and M.~Rush.
%\newblock Enzyme-catalyzed cellular transanimations.
%\newblock In A.F. Round, editor, \emph{Advances in Enzymology}, volume~2, pages
%  125--247. Academic Press, New York, 3rd edition, 1954.

%\bibitem[Keohane(1958)]{Keo:58}
%\newblock \emph{Power and Interdependence: World Politics in Transitions}.
%\newblock Little, Brown \& Co., Boston, 1958.

%\bibitem[Powers(1985)]{Pow:85}
%\newblock Is there a way out?
%\newblock \emph{Harpers}, pages 35--47, June 1985.

%\bibitem[Soukhanov(1992)]{Heritage:92}
%\newblock \emph{{The American Heritage. Dictionary of the American Language}}.
%\newblock Houghton Mifflin Company, 1992.

%\end{thebibliography}
                                                                         % in the appendices.
\end{document}